\documentclass[5p,twocolumn]{elsarticle}

\usepackage{lineno,hyperref}
\modulolinenumbers[5]

\usepackage{amsmath}
\usepackage{amssymb}

\usepackage{graphicx}
\usepackage{float}
\usepackage{xcolor}
\usepackage[english]{babel}
\usepackage[output-decimal-marker={.},exponent-product=\cdot]{siunitx}

\biboptions{sort&compress}

\hypersetup{colorlinks,allcolors=black}
\journal{The European Physical Journal A}

\begin{document}

\begin{frontmatter}

\title{Reinvestigation of $^{91}$Sr and $^{95}$Y atomic masses using the JYFLTRAP Penning trap}

\author[1]{A.~Jaries\corref{cor}}
\ead{arthur.a.jaries@jyu.fi}
\author[1]{M.~Stryjczyk\corref{cor}}
\ead{marek.m.stryjczyk@jyu.fi}
\author[1]{A.~Kankainen\corref{cor}}
\ead{anu.kankainen@jyu.fi}
\author[1]{T.~Eronen}
\author[1]{Z.~Ge}
\author[1]{M.~Mougeot}
\author[1]{A.~Raggio}
\author[1]{J.~Ruotsalainen}

\affiliation[1]{organization={University of Jyvaskyla, Department of Physics, Accelerator laboratory},
addressline={P.O. Box 35(YFL)},
postcode={FI-40014},
city={University of Jyvaskyla},
country={Finland}}

\cortext[cor]{Corresponding authors}

\begin{abstract}
We report on the precise mass measurements of the $^{91}$Sr and $^{95}$Y isotopes performed using the JYFLTRAP double Penning trap mass spectrometer. The mass-excess values from this work, ${\mathrm{ME}(^{91}\mathrm{Sr}) = -83645.5(13)}$~keV and ${\mathrm{ME}(^{95}\mathrm{Y}) = -81226.4(10)}$~keV, deviate by 6.5(52)~keV and $-18(7)$~keV from the Atomic Mass Evaluation 2020 (AME20). In the case of $^{91}$Sr the new result disagrees with the ISOLTRAP value, while for $^{95}$Y, it agrees with the older JYFLTRAP value.
\end{abstract}

\begin{keyword}
Binding energies and masses \sep Mass spectrometers \sep Penning trap
\end{keyword}

\end{frontmatter}


\section{Introduction}

The mass is one of the most fundamental properties of a nucleus as it is a reflection of all the interactions between the constituent nucleons. In addition, changes in mass trends along an isotopic or isotonic chain can reveal information about the structure of the ground states \cite{Lunney2003,Eronen2016}. Masses are necessary for accessing other experimental information, such as the determination of $\log(ft)$ values in $\beta$-decay spectroscopy \cite{Turkat2023} or differences in charge radii in laser spectroscopy \cite{Yang2023}. They also have an influence on astrophysical calculations, for instance the r-process abundance predictions \cite{Mumpower2016}. 

Because of the influence of masses on nuclear physics, a review of available data, the Atomic Mass Evaluation (AME) \cite{AME2020}, is prepared periodically, most recently in 2020, where all pieces of information which can be used for mass determination are summarized and critically evaluated. The AME also points to discrepancies in the literature and, if needed, rejects data points deemed unreliable.

There are several experimental approaches which enable the extraction of atomic masses, see Ref. \cite{Huang2021} for an overview. The method which provides the best accuracy and resolving power is the Penning-trap mass spectrometry \cite{Eronen2016,Dilling2018,Huang2021}. With the recent advent of the phase-imaging ion-cyclotron-resonance (PI-ICR) technique \cite{Eliseev2013,Eliseev2014,Nesterenko2018}, the improvement in resolving power is such that states as close as 10 keV can now be separated \cite{Huang2021,Orford2018}.

Because of its high reliability, the rejection of the Penning-trap measurements in favor of results from other experimental techniques is not common. Nevertheless, it is the case for $^{91}$Sr and $^{95}$Y \cite{Huang2021}. While their masses were measured with Penning traps using the Time-of-Flight Ion Cyclotron Resonance (TOF-ICR) technique \cite{Konig1995,Graff1980} at ISOLTRAP \cite{Raimbault-Hartmann2002} and JYFLTRAP \cite{Hager2007}, respectively, the reported values were rejected from the AME and the results were labeled as 'Well-documented data, or data from regular reviewed journals, which disagree with other well-documented values.' \cite{Huang2021}. 

Currently, the mass of $^{91}$Sr is determined exclusively from decay measurements \cite{AME2020}, mostly the $\beta$ decay of $^{91}$Sr to $^{91}$Y (81\%) \cite{Halbig1973,Decker1980,Iafigliola1983}, with the remaining 12\% from the $\beta$ and $\beta$-delayed-neutron studies of $^{91,92}$Rb \cite{Decker1980,Iafigliola1983,Przewloka1992,Kratz1984}. At the same time, the mass of $^{95}$Y is extracted about 88\% from $\beta$ decays \cite{AME2020} (56\% from $^{95}\mathrm{Y}\rightarrow^{95}$Zr \cite{Decker1980}, 32\% from $^{95}\mathrm{Sr}\rightarrow^{95}$Y \cite{Bloennigen1984}) and 12\% from the $^{96}\mathrm{Zr}(t,\alpha)^{95}\mathrm{Y}$ transfer reaction \cite{AME2020,Flynn1983}. While $\beta$-decay studies are known to be unreliable, especially for nuclei far from stability where measurements can suffer from the pandemonium effect \cite{Audi2012,Hardy1977}, for $^{91}$Sr and $^{95}$Y there are several measurements agreeing with each other but differing from the Penning-trap values by 3.0 and 1.8 standard deviations ($\sigma$), respectively \cite{Huang2021}. 

The neutron-rich nuclei in the $A=90$ region are abundantly produced in fission and, as a result, they contribute to the decay heat generated in nuclear reactors \cite{Nichols2023}. In addition, three reactions, $^{88}$Kr$(\alpha,n)^{91}$Sr, $^{91}$Sr$(\alpha,n)^{94}$Zr and $^{95}$Y$(\alpha,n)^{98}$Nb, were identified to play an important role in the production of lighter heavy elements between Sr and Ag in neutrino-driven, neutron-rich ejecta of core-collapse supernovae \cite{Bliss2020}. Thus, it is important to have reliable mass values for the nuclei of interest. 

To resolve the discrepancy in existing literature, in this work we report on the results of the $^{91}$Sr and $^{95}$Y \mbox{PI-ICR} mass measurements performed using the JYFLTRAP double Penning trap mass spectrometer \cite{Eronen2012} at the Ion Guide Isotope Separator On-Line (IGISOL) facility \cite{Moore2013,Penttila2020} in the JYFL Accelerator Laboratory at the University of Jyv\"askyl\"a, Finland.

\section{Experimental method and results}

The radioactive species were produced in a proton-induced fission of a 15~mg/cm$^2$ thick $^{nat}$U target by impinging a 25-MeV primary proton beam delivered by the K130 cyclotron. The primary beam current was about 1~$\mu$A to produce $^{91}$Sr and about 5~$\mu$A for $^{95}$Y. The fission products were first stopped in a gas cell filled with helium gas at a pressure of about 280~mbar from which they were extracted following gas flow and guided to the high-vacuum region of the mass separator using a sextupole ion guide \cite{Karvonen2008}. Subsequently, the ions were accelerated to 30$q$~kV energy. The beam was separated with respect to their mass-over-charge ratio by a 55$^{\circ}$ dipole magnet with a mass resolving power of $m/\Delta m \approx 500$ and injected into a radio-frequency quadrupole cooler-buncher \cite{Nieminen2001}. From there, the cooled and bunched radioactive ion beam was finally delivered to the JYFLTRAP double Penning trap \cite{Eronen2012}. 

\begin{figure}[h!t!b]
\includegraphics[width=\columnwidth]{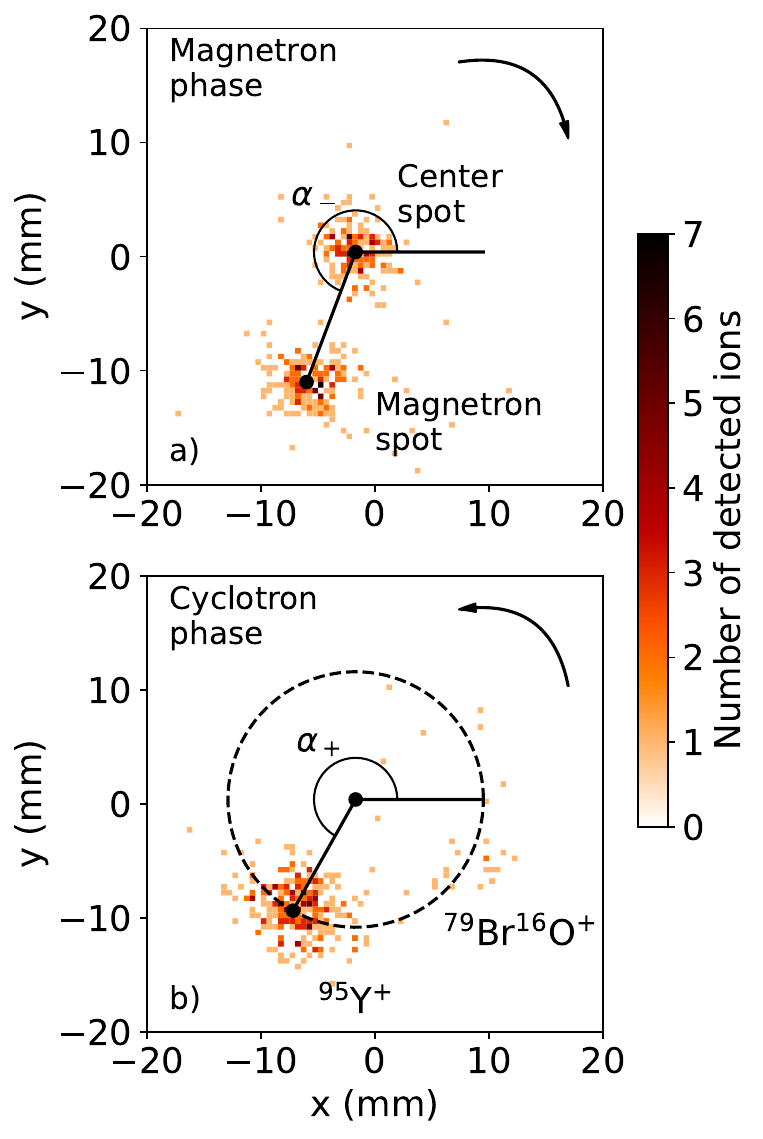}
\caption{\label{fig:95YPIICR} Projection of \textbf{a)} the magnetron motion and \textbf{b)} the cyclotron motion of $^{95}$Y$^+$ and the molecular contaminant $^{79}$Br$^{16}$O$^+$ ions onto the position-sensitive detector obtained with the PI-ICR technique using a phase accumulation time $t_{acc} = 694$~ms. The center spot is shown on panel \textbf{a)}. The average excitation radius is indicated with the dashed circle on panel \textbf{b)}. The center positions of the spots are indicated with the dot symbol. The angular positions $\alpha_-$ and $\alpha_+$ of the respective magnetron and cyclotron phase projections are marked with an arc.}
\end{figure}

At JYFLTRAP, the ions were first cooled, purified and centered in the first (preparation) trap by using a mass-selective buffer-gas cooling technique \cite{Savard1991}. A mass resolving power of $m/\Delta m > 10^4$ was reached which enabled removal of the vast majority of isobaric contaminants. They were subsequently transferred to the second (measurement) trap through a 1.5-mm diameter diaphragm and, after about 600 $\mu$s, the purified ions of interest were transferred back to the purifying trap for additional cooling. Finally, the singly-charged ions of interest were sent to the measurement trap, where their cyclotron frequency $\nu_c = qB/(2 \pi m)$ in the magnetic field $B$ was determined using the PI-ICR technique \cite{Eliseev2013,Eliseev2014,Nesterenko2018}. In this technique, the cyclotron frequency of an ion is obtained from the angular difference $\alpha_c = \alpha_+ - \alpha_-$ between the projections of the cyclotron ($\alpha_+$) and magnetron ($\alpha_-$) radial in-trap motion images, see Fig. \ref{fig:95YPIICR}. They are measured on the position-sensitive detector with respect to the trap center during a phase accumulation time $t_{acc}$. In the present case, $t_{acc}$ was set to 627.4~ms for $^{91}$Sr while for $^{95}$Y two measurements were performed, with 694 and 713~ms accumulation times.

The magnetic field strength $B$ was determined using $^{85}$Rb reference ions (mass excess ${\mathrm{ME}_{lit.} = -82167.341(5)}$~keV \cite{AME2020}) produced by an offline surface ion source \cite{Vilen2020}. The atomic mass $M$ is determined from the cyclotron frequency ratio $r=\nu_{c,ref}/\nu_{c}$ of the reference ion and the ion of interest:
\begin{equation}
M = (M_{ref} - m_{e}) r + m_{e}\mathrm{,}
\end{equation}
where $M_{ref}$ and $m_{e}$ are the atomic mass of the reference ion the electron mass, respectively. As the binding energy of the missing electron is of the order of a few eV, its contribution was neglected. 

The measurements of the ion of interest and the reference ion were done alternately to reduce systematic effects due to magnetic field fluctuations. To assess the ion-ion interactions in the measurement trap, the count-rate class analysis \cite{Roux2013} was performed for $^{95}$Y, however, no significant dependency was observed. For $^{91}$Sr this analysis was not statistically feasible, thus, the data was limited to one detected ion per bunch. A mass-dependent uncertainty of $\delta r/r = -2.35(81) \times 10^{-10} / \textnormal{u} \times (M_{ref} - M)$ and a residual systematic uncertainty of $\delta r/r=9\times 10^{-9}$ were added to the cyclotron frequency ratio \cite{Nesterenko2021}. In addition, the systematic uncertainties due to the temporal magnetic field fluctuation ($\delta B/B = 2.01(25) \times 10^{-12}$ min$^{-1} \times \delta t$, where $\delta t$ is the time between the measurements), the magnetron phase advancement and the angle error were also included in the uncertainty estimation \cite{Nesterenko2021}. 

\begin{table*}[h!t!b]
\centering\small
\caption{\label{tab:results} Frequency ratios $r=\nu_{c,ref}/\nu_{c}$ determined in this work using the PI-ICR technique with a given accumulation time $t_{acc}$ and their corresponding mass-excess values (ME). In both cases $^{85}$Rb (ME$_\mathrm{lit.} = -82167.341(5)$~keV \cite{AME2020}) was used as a reference ion. For comparison, mass-excess values from AME20 (ME$_\mathrm{AME}$) \cite{AME2020} and from the Penning-trap measurements (ME$_\mathrm{trap}$) \cite{Raimbault-Hartmann2002,Hager2007}, which are recalculated using reported frequency ratios $r$ and masses of reference isotopes from AME20 \cite{AME2020}, are presented. In addition, differences between this work and AME20 (${\mathrm{Diff}_\mathrm{AME} = \mathrm{ME}-\mathrm{ME}_\mathrm{AME}}$) and between this work and Penning-trap measurements (${\mathrm{Diff}_\mathrm{trap} = \mathrm{ME}-\mathrm{ME}_\mathrm{trap}}$) are also listed. Half-lives $T_{1/2}$ and spin-parity assignments $J^{\pi}$ of the measured species  are taken from the literature \cite{NUBASE2020}.}
\begin{tabular}{llllllllll}
\hline\noalign{\smallskip}
Nuclide & $T_{1/2}$ & $J^{\pi}$ & $t_{acc}$ & $r=\nu_{c,ref}/\nu_{c}$ & ME & ME$_\mathrm{AME}$ & Diff$_\mathrm{AME}$ & ME$_\mathrm{trap}$ & Diff$_\mathrm{trap}$ \\
 & & & (ms) & & (keV) & (keV) & (keV) & (keV) & (keV) \\\hline
$^{91}$Sr & 9.65(6) h & $5/2^+$ & 627.4 & \num{1.070643334(13)} & \num{-83645.5(13)} & \num{-83652(5)} & \num{6.5(52)} & \num{-83621(9)} & \num{-25(9)} \\
$^{95}$Y & 10.3(1) m & $1/2^-$ & 694, 713 & \num{1.117781933(13)} & \num{-81226.4(10)} & \num{-81208(7)} & \num{-18(7)} & \num{-81225(6)} & \num{-1.4(61)} \\\hline\noalign{\smallskip}
\end{tabular}
\end{table*}

\begin{figure*}[h!t!b]
\centering
\includegraphics[width=\textwidth]{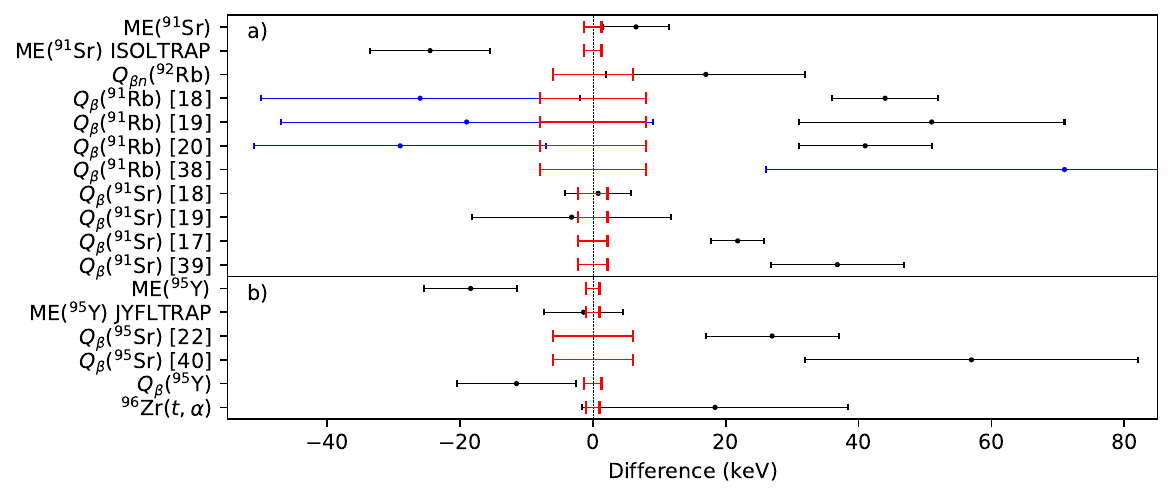}
\caption{\label{fig:comparison}Differences of mass-related properties between this work (red) and the literature (black) for a) $^{91}$Sr and b) $^{95}$Y, defined as $\mathrm{Difference} = \mathrm{This~work} - \mathrm{Literature}$. Mass-excess values of $^{91}$Sr and $^{95}$Y from AME20 \cite{AME2020} [ME($^{91}$Sr), ME($^{95}$Y)] and from the previous Penning-trap measurements [ME($^{91}$Sr)~ISOLTRAP, ME($^{95}$Y)~JYFLTRAP] \cite{Raimbault-Hartmann2002,Hager2007} are presented. The $\beta$-delayed-neutron energy of $^{92}$Rb [$Q_{\beta n}(^{92}\mathrm{Rb})$] is from Ref. \cite{Kratz1984}, the $\beta$-decay energies of $^{91}$Rb [$Q_\beta(^{91}\mathrm{Rb})$], $^{91}$Sr [$Q_\beta(^{91}\mathrm{Sr})$], $^{95}$Sr [$Q_\beta(^{95}\mathrm{Sr})$] and $^{95}$Y [$Q_\beta(^{95}\mathrm{Y})$] are from Refs. \cite{Decker1980,Iafigliola1983,Przewloka1992,Wohn1978,Halbig1973,Ames1953,Bloennigen1984,Mach1990} and the $Q$ value of the $^{96}\mathrm{Zr}(t,\alpha)$ reaction [$^{96}\mathrm{Zr}(t,\alpha)$] is from Ref. \cite{Flynn1983}. Blue bars for $Q_\beta(^{91}\mathrm{Rb})$ indicate values adjusted by the AME20 evaluators. Only the adjusted $Q_\beta(^{91}\mathrm{Rb})$ value from \cite{Wohn1978} is shown.}
\end{figure*}

A summary of experimental results as well as a comparison with AME20 \cite{AME2020} and the rejected Penning-trap values \cite{Raimbault-Hartmann2002,Hager2007} is presented in Table~\ref{tab:results}. We note that the latter were recalculated using the reported frequency ratios $r$ and the masses of the reference isotopes from AME20 \cite{AME2020}. In addition, a comparison between this work and input data used in AME20 is shown in Fig. \ref{fig:comparison}.

The mass-excess value of $^{91}$Sr from this work, ${\mathrm{ME} = -83645.5(13)}$~keV differs by $6.5(52)$~keV (1.3$\sigma$) from the AME20 value \cite{AME2020} but it is four times more precise. At the same time it is $-25(9)$~keV (2.7$\sigma$) away from the ISOLTRAP result \cite{Raimbault-Hartmann2002}, indicating that the decision of the AME20 evaluators to exclude this data point was correct. The exact reason why this value is incorrect remains unknown. However, we note that Ref. \cite{Raimbault-Hartmann2002}  is one of the earlier publications from ISOLTRAP. At the time, the systematic effects had not yet been studied in detail but published later in Ref. \cite{Kellerbauer2003}. In addition, the preparation trap consisted of a 0.7 T electromagnet and its limited resolving power could have led to a presence of the $^{91}$Rb contamination. This hypothesis is supported by the fact that the ISOLTRAP value is shifted towards heavier masses. 

The updated $Q_{\beta n}(^{92}\mathrm{Rb})$ value from this work, 802(6)~keV, is closer to 785(15)~keV reported in Ref. \cite{Kratz1984} compared to 808(8)~keV from AME20 \cite{AME2020}. The new $Q_\beta(^{91}\mathrm{Rb})$ value, 5901(8)~keV, is larger than the three results from the $\beta$-decay studies taken into account in AME20: 5857(8)~keV from Ref.~\cite{Decker1980}, 5850(20)~keV from Ref.~\cite{Iafigliola1983} and 5860(10)~keV from Ref.~\cite{Przewloka1992}, see Fig.~\ref{fig:comparison}a. However, these data points were adjusted by the evaluators \cite{Huang2021}, as indicated with blue bars in Fig.~\ref{fig:comparison}a, to include the fact that the 94-keV state in $^{91}$Sr is fed significantly stronger than the ground state in the $\beta$ decay of $^{91}$Rb \cite{Baglin2013,Huang2021}. A deviation between the results from this work and the AME20-adjusted values might be related to the fact that the $\beta$ feedings to the two low-lying states are actually similar, as indicated by the recent total absorption spectroscopy study \cite{Rice2017}. We note that the $Q_\beta(^{91}\mathrm{Rb})$ value from Ref.~\cite{Wohn1978}, 5760(40)~keV, was not included in the AME20 evaluation due to a large uncertainty. 

The updated $Q_\beta(^{91}\mathrm{Sr})$ value, 2705.8(22)~keV, agrees very well with the results of two $\beta$-decay studies, 2705(5)~keV reported in Ref. \cite{Decker1980} and 2709(15)~keV from Ref. \cite{Iafigliola1983}. However, it disagrees with 2684(4)~keV and 2665(10)~keV from Refs. \cite{Halbig1973,Ames1953} by 4.8 and 4.0$\sigma$, respectively, see Fig. \ref{fig:comparison}a. We note that the uncertainty of the value reported in Ref. \cite{Halbig1973} was increased by the AME20 evaluators from 4 to 10 keV \cite{Huang2021} while the result from Ref.~\cite{Ames1953} was rejected.  

The mass-excess value of $^{95}$Y from this work, ${\mathrm{ME} = -81226.4(10)}$~keV differs by $-18(7)$~keV (2.6$\sigma$) from AME20 \cite{AME2020}. However, it is in a perfect agreement with the previous JYFLTRAP measurement performed using the TOF-ICR technique and $^{97}$Zr as a reference nucleus \cite{Hager2007} and it is four times more precise. During the measurement a second cyclotron spot was observed, see Fig. \ref{fig:95YPIICR}. The extracted mass-excess value, ${\mathrm{ME} = -80804.9(12)}$~keV, allowed us to identify it as an isobaric molecular contamination of $^{79}$Br$^{16}$O (${\mathrm{ME}_\mathrm{lit.} = -80805.1(10)}$~keV \cite{AME2020}).

The updated $Q_\beta(^{95}\mathrm{Sr})$ value, 6109(6)~keV, is 27~keV (2.3$\sigma$) larger than 6082(10)~keV reported in Ref. \cite{Bloennigen1984}. By discarding it, the consistency in the region can be restored. This is justified since this result 'should be considered preliminary, until the detector response is evaluated in detail', as it is stated in Ref.~\cite{Bloennigen1984}. We note that our $Q_\beta(^{95}\mathrm{Sr})$ value differs by more than $2\sigma$ from 6052(25)~keV reported in Ref.~\cite{Mach1990}, which was not taken into account in AME20 due to a large uncertainty \cite{Huang2021}. The $Q_\beta(^{95}\mathrm{Y})$ value from Ref.~\cite{Decker1980} (4445(9)~keV) and the $Q$ value of the $^{96}\mathrm{Zr}(t,\alpha)$ reaction from Ref. \cite{Flynn1983} (8294(20)~keV) agree relatively well with the updated results from this work (4433.5(13)~keV and 8312.4(10)~keV, respectively), as can be seen in Fig.~\ref{fig:comparison}b. 

\section{Conclusions}

Masses of $^{91}$Sr and $^{95}$Y were measured using the JYFLTRAP double Penning trap. The extracted mass-excess value of $^{91}$Sr agrees with AME20 but it does not match the previous ISOLTRAP measurement. For $^{95}$Y the mass-excess value differs by $-18(7)$~keV from AME20 but it perfectly agrees with the previous JYFLTRAP measurement. Our study shows an importance of critical mass evaluation and cross checks of different experimental results.

\section*{Acknowledgments}

This project has received funding from the European Union’s Horizon 2020 research and innovation programme under Grant Agreements No. 771036 (ERC CoG MAIDEN) and No. 861198–LISA–H2020-MSCA-ITN-2019, from the European Union’s Horizon Europe Research and Innovation Programme under Grant Agreement No. 101057511 (EURO-LABS) and from the Academy of Finland projects No. 295207, 306980, 327629, 354589 and 354968. J.R. acknowledges financial support from the Vilho, Yrj\"o and Kalle V\"ais\"al\"a Foundation.

\bibliography{mybibfile}

\begin{thebibliography}{10}
\expandafter\ifx\csname url\endcsname\relax
  \def\url#1{\texttt{#1}}\fi
\expandafter\ifx\csname urlprefix\endcsname\relax\def\urlprefix{URL }\fi
\expandafter\ifx\csname href\endcsname\relax
  \def\href#1#2{#2} \def\path#1{#1}\fi

\bibitem{Lunney2003}
D.~Lunney, J.~M. Pearson, C.~Thibault, \href{https://link.aps.org/doi/10.1103/RevModPhys.75.1021}{{Recent trends in the determination of nuclear masses}}, Rev. Mod. Phys. 75 (2003) 1021--1082.
\newblock \href {https://doi.org/10.1103/RevModPhys.75.1021} {\path{doi:10.1103/RevModPhys.75.1021}}.
\newline\urlprefix\url{https://link.aps.org/doi/10.1103/RevModPhys.75.1021}

\bibitem{Eronen2016}
T.~Eronen, A.~Kankainen, J.~Äystö, \href{https://www.sciencedirect.com/science/article/pii/S0146641016300436}{{Ion traps in nuclear physics - Recent results and achievements}}, Progress in Particle and Nuclear Physics 91 (2016) 259--293.
\newblock \href {https://doi.org/10.1016/j.ppnp.2016.08.001} {\path{doi:10.1016/j.ppnp.2016.08.001}}.
\newline\urlprefix\url{https://www.sciencedirect.com/science/article/pii/S0146641016300436}

\bibitem{Turkat2023}
S.~Turkat, X.~Mougeot, B.~Singh, K.~Zuber, \href{https://www.sciencedirect.com/science/article/pii/S0092640X23000128}{{Systematics of logft values for $\beta^-$, and EC/$\beta^+$ transitions}}, Atomic Data and Nuclear Data Tables 152 (2023) 101584.
\newblock \href {https://doi.org/10.1016/j.adt.2023.101584} {\path{doi:10.1016/j.adt.2023.101584}}.
\newline\urlprefix\url{https://www.sciencedirect.com/science/article/pii/S0092640X23000128}

\bibitem{Yang2023}
X.~Yang, S.~Wang, S.~Wilkins, R.~G. Ruiz, \href{https://www.sciencedirect.com/science/article/pii/S0146641022000631}{{Laser spectroscopy for the study of exotic nuclei}}, Progress in Particle and Nuclear Physics 129 (2023) 104005.
\newblock \href {https://doi.org/10.1016/j.ppnp.2022.104005} {\path{doi:10.1016/j.ppnp.2022.104005}}.
\newline\urlprefix\url{https://www.sciencedirect.com/science/article/pii/S0146641022000631}

\bibitem{Mumpower2016}
M.~Mumpower, R.~Surman, G.~McLaughlin, A.~Aprahamian, \href{https://www.sciencedirect.com/science/article/pii/S0146641015000897}{{The impact of individual nuclear properties on r-process nucleosynthesis}}, Progress in Particle and Nuclear Physics 86 (2016) 86--126.
\newblock \href {https://doi.org/10.1016/j.ppnp.2015.09.001} {\path{doi:10.1016/j.ppnp.2015.09.001}}.
\newline\urlprefix\url{https://www.sciencedirect.com/science/article/pii/S0146641015000897}

\bibitem{AME2020}
M.~Wang, W.~Huang, F.~Kondev, G.~Audi, S.~Naimi, \href{https://doi.org/10.1088/1674-1137/abddaf}{{The {AME} 2020 atomic mass evaluation ({II}). {T}ables, graphs and references}}, Chinese Physics C 45~(3) (2021) 030003.
\newblock \href {https://doi.org/10.1088/1674-1137/abddaf} {\path{doi:10.1088/1674-1137/abddaf}}.
\newline\urlprefix\url{https://doi.org/10.1088/1674-1137/abddaf}

\bibitem{Huang2021}
W.~Huang, M.~Wang, F.~Kondev, G.~Audi, S.~Naimi, \href{https://doi.org/10.1088/1674-1137/abddb0}{{The AME 2020 atomic mass evaluation (I). Evaluation of input data, and adjustment procedures}}, Chinese Physics C 45~(3) (2021) 030002.
\newblock \href {https://doi.org/10.1088/1674-1137/abddb0} {\path{doi:10.1088/1674-1137/abddb0}}.
\newline\urlprefix\url{https://doi.org/10.1088/1674-1137/abddb0}

\bibitem{Dilling2018}
J.~Dilling, K.~Blaum, M.~Brodeur, S.~Eliseev, \href{https://doi.org/10.1146/annurev-nucl-102711-094939}{{Penning-Trap Mass Measurements in Atomic and Nuclear Physics}}, Annual Review of Nuclear and Particle Science 68~(1) (2018) 45--74.
\newblock \href {https://doi.org/10.1146/annurev-nucl-102711-094939} {\path{doi:10.1146/annurev-nucl-102711-094939}}.
\newline\urlprefix\url{https://doi.org/10.1146/annurev-nucl-102711-094939}

\bibitem{Eliseev2013}
S.~Eliseev, K.~Blaum, M.~Block, C.~Droese, M.~Goncharov, E.~Minaya~Ramirez, D.~A. Nesterenko, Y.~N. Novikov, L.~Schweikhard, \href{https://link.aps.org/doi/10.1103/PhysRevLett.110.082501}{{Phase-{I}maging {I}on-{C}yclotron-{R}esonance {M}easurements for {S}hort-{L}ived {N}uclides}}, Phys. Rev. Lett. 110 (2013) 082501.
\newblock \href {https://doi.org/10.1103/PhysRevLett.110.082501} {\path{doi:10.1103/PhysRevLett.110.082501}}.
\newline\urlprefix\url{https://link.aps.org/doi/10.1103/PhysRevLett.110.082501}

\bibitem{Eliseev2014}
S.~Eliseev, K.~Blaum, M.~Block, A.~D{\"o}rr, C.~Droese, T.~Eronen, M.~Goncharov, M.~H{\"o}cker, J.~Ketter, E.~M. Ramirez, D.~A. Nesterenko, Y.~N. Novikov, L.~Schweikhard, \href{https://doi.org/10.1007/s00340-013-5621-0}{{A phase-imaging technique for cyclotron-frequency measurements}}, Applied Physics B 114~(1) (2014) 107--128.
\newblock \href {https://doi.org/10.1007/s00340-013-5621-0} {\path{doi:10.1007/s00340-013-5621-0}}.
\newline\urlprefix\url{https://doi.org/10.1007/s00340-013-5621-0}

\bibitem{Nesterenko2018}
D.~A. Nesterenko, T.~Eronen, A.~Kankainen, L.~Canete, A.~Jokinen, I.~D. Moore, H.~Penttil{\"a}, S.~Rinta-Antila, A.~de~Roubin, M.~Vilen, \href{https://doi.org/10.1140/epja/i2018-12589-y}{{Phase-{I}maging {I}on-{C}yclotron-{R}esonance technique at the {JYFLTRAP} double {P}enning trap mass spectrometer}}, The European Physical Journal A 54~(9) (2018) 154.
\newblock \href {https://doi.org/10.1140/epja/i2018-12589-y} {\path{doi:10.1140/epja/i2018-12589-y}}.
\newline\urlprefix\url{https://doi.org/10.1140/epja/i2018-12589-y}

\bibitem{Orford2018}
R.~Orford, \href{https://escholarship.mcgill.ca/concern/theses/6h440v95x}{{A phase-imaging technique for precision mass measurements of neutron-rich nuclei with the {C}anadian {P}enning {T}rap mass spectrometer}}, Ph.D. thesis, McGill University (2018).
\newline\urlprefix\url{https://escholarship.mcgill.ca/concern/theses/6h440v95x}

\bibitem{Konig1995}
M.~K{\"o}nig, G.~Bollen, H.-J. Kluge, T.~Otto, J.~Szerypo, \href{https://www.sciencedirect.com/science/article/pii/016811769504146C}{{Quadrupole excitation of stored ion motion at the true cyclotron frequency}}, International Journal of Mass Spectrometry and Ion Processes 142~(1) (1995) 95--116.
\newblock \href {https://doi.org/10.1016/0168-1176(95)04146-C} {\path{doi:10.1016/0168-1176(95)04146-C}}.
\newline\urlprefix\url{https://www.sciencedirect.com/science/article/pii/016811769504146C}

\bibitem{Graff1980}
G.~Gr{\"a}ff, H.~Kalinowsky, J.~Traut, \href{https://doi.org/10.1007/BF01414243}{{A direct determination of the proton electron mass ratio}}, Zeitschrift f{\"u}r Physik A Atoms and Nuclei 297~(1) (1980) 35--39.
\newblock \href {https://doi.org/10.1007/BF01414243} {\path{doi:10.1007/BF01414243}}.
\newline\urlprefix\url{https://doi.org/10.1007/BF01414243}

\bibitem{Raimbault-Hartmann2002}
H.~Raimbault-Hartmann, G.~Audi, D.~Beck, G.~Bollen, M.~{de Saint Simon}, H.-J. Kluge, M.~König, R.~Moore, S.~Schwarz, G.~Savard, J.~Szerypo, \href{https://www.sciencedirect.com/science/article/pii/S0375947402008631}{{High-accuracy mass determination of neutron-rich rubidium and strontium isotopes}}, Nuclear Physics A 706~(1) (2002) 3--14.
\newblock \href {https://doi.org/10.1016/S0375-9474(02)00863-1} {\path{doi:10.1016/S0375-9474(02)00863-1}}.
\newline\urlprefix\url{https://www.sciencedirect.com/science/article/pii/S0375947402008631}

\bibitem{Hager2007}
U.~Hager, A.~Jokinen, V.-V. Elomaa, T.~Eronen, J.~Hakala, A.~Kankainen, S.~Rahaman, J.~Rissanen, I.~Moore, S.~Rinta-Antila, A.~Saastamoinen, T.~Sonoda, J.~Äystö, \href{https://www.sciencedirect.com/science/article/pii/S0375947407006070}{{Precision mass measurements of neutron-rich yttrium and niobium isotopes}}, Nuclear Physics A 793~(1) (2007) 20--39.
\newblock \href {https://doi.org/10.1016/j.nuclphysa.2007.06.011} {\path{doi:10.1016/j.nuclphysa.2007.06.011}}.
\newline\urlprefix\url{https://www.sciencedirect.com/science/article/pii/S0375947407006070}

\bibitem{Halbig1973}
J.~Halbig, F.~Wohn, W.~Talbert, J.~Eitter, J.~McConnell, \href{https://www.sciencedirect.com/science/article/pii/037594747390362X}{{The $\beta$-decay of 91Sr}}, Nuclear Physics A 203~(3) (1973) 532--538.
\newblock \href {https://doi.org/10.1016/0375-9474(73)90362-X} {\path{doi:10.1016/0375-9474(73)90362-X}}.
\newline\urlprefix\url{https://www.sciencedirect.com/science/article/pii/037594747390362X}

\bibitem{Decker1980}
R.~Decker, K.~D. W{\"u}nsch, H.~Wollnik, E.~Koglin, G.~Siegert, G.~Jung, \href{https://doi.org/10.1007/BF01473120}{{Pr{\"a}zise $Q_\beta$-Wert-Messungen mit einem Intrinsic-Germanium Detektor an leichten, neutronenreichen Spaltprodukten}}, Zeitschrift f{\"u}r Physik A Atoms and Nuclei 294~(1) (1980) 35--49.
\newblock \href {https://doi.org/10.1007/BF01473120} {\path{doi:10.1007/BF01473120}}.
\newline\urlprefix\url{https://doi.org/10.1007/BF01473120}

\bibitem{Iafigliola1983}
R.~Iafigliola, M.~Chatterjee, H.~Dautet, J.~K.~P. Lee, \href{https://doi.org/10.1139/v83-128}{{Precise Q$_\beta$ measurements for A=91 to 93 mass chains}}, Canadian Journal of Chemistry 61~(4) (1983) 694--696.
\newblock \href {https://doi.org/10.1139/v83-128} {\path{doi:10.1139/v83-128}}.
\newline\urlprefix\url{https://doi.org/10.1139/v83-128}

\bibitem{Przewloka1992}
M.~Przewloka, A.~Przewloka, P.~W{\"a}chter, H.~Wollnik, \href{https://doi.org/10.1007/BF01294483}{{Measurements of $\beta$-endpoint-energies using a magnetic electron separator}}, Zeitschrift f{\"u}r Physik A Hadrons and Nuclei 342~(1) (1992) 23--26.
\newblock \href {https://doi.org/10.1007/BF01294483} {\path{doi:10.1007/BF01294483}}.
\newline\urlprefix\url{https://doi.org/10.1007/BF01294483}

\bibitem{Kratz1984}
K.~Kratz, A.~Schroeder, H.~Ohm, H.~Gabelmann, W.~Ziegert, B.~Steinmueller, B.~Pfeiffer, \href{http://inis.iaea.org/search/search.aspx?orig_q=RN:16044406}{{Determination of Q$_\beta$-B$_n$ values of neutron-rich A=90 and 140 nuclei}}, in: Atomic masses and fundamental constants, 1984, pp. 127--133.
\newline\urlprefix\url{http://inis.iaea.org/search/search.aspx?orig_q=RN:16044406}

\bibitem{Bloennigen1984}
F.~Bloennigen, B.~Pfeiffer, G.~Bewersdorf, C.~Geisse, W.~Lippert, U.~Stoehlker, H.~Wollnik, \href{http://inis.iaea.org/search/search.aspx?orig_q=RN:16044423}{{Precise $Q_\beta$-measurements for A=95, 96, 97, 98 mass chains with a GE(HP)-detector-telescope}}, in: Atomic masses and fundamental constants, 1984, pp. 134--140.
\newline\urlprefix\url{http://inis.iaea.org/search/search.aspx?orig_q=RN:16044423}

\bibitem{Flynn1983}
E.~R. Flynn, R.~E. Brown, F.~Ajzenberg-Selove, J.~A. Cizewski, \href{https://link.aps.org/doi/10.1103/PhysRevC.28.575}{{Proton hole states in $^{95}\mathrm{Y}$ and $^{95,97,99}\mathrm{Nb}$}}, Phys. Rev. C 28 (1983) 575--586.
\newblock \href {https://doi.org/10.1103/PhysRevC.28.575} {\path{doi:10.1103/PhysRevC.28.575}}.
\newline\urlprefix\url{https://link.aps.org/doi/10.1103/PhysRevC.28.575}

\bibitem{Audi2012}
G.~Audi, M.~Wang, A.~Wapstra, F.~Kondev, M.~MacCormick, X.~Xu, B.~Pfeiffer, \href{https://doi.org/10.1088/1674-1137/36/12/002}{{The Ame2012 atomic mass evaluation}}, Chinese Physics C 36~(12) (2012) 1287.
\newblock \href {https://doi.org/10.1088/1674-1137/36/12/002} {\path{doi:10.1088/1674-1137/36/12/002}}.
\newline\urlprefix\url{https://doi.org/10.1088/1674-1137/36/12/002}

\bibitem{Hardy1977}
J.~Hardy, L.~Carraz, B.~Jonson, P.~Hansen, \href{https://www.sciencedirect.com/science/article/pii/0370269377902234}{{The essential decay of pandemonium: A demonstration of errors in complex beta-decay schemes}}, Physics Letters B 71~(2) (1977) 307--310.
\newblock \href {https://doi.org/10.1016/0370-2693(77)90223-4} {\path{doi:10.1016/0370-2693(77)90223-4}}.
\newline\urlprefix\url{https://www.sciencedirect.com/science/article/pii/0370269377902234}

\bibitem{Nichols2023}
A.~L. Nichols, P.~Dimitriou, A.~Algora, M.~Fallot, L.~Giot, F.~G. Kondev, T.~Yoshida, M.~Karny, G.~Mukherjee, B.~C. Rasco, K.~P. Rykaczewski, A.~A. Sonzogni, J.~L. Tain, \href{https://doi.org/10.1140/epja/s10050-023-00969-x}{{Improving fission-product decay data for reactor applications: part I---decay heat}}, The European Physical Journal A 59~(4) (2023) 78.
\newblock \href {https://doi.org/10.1140/epja/s10050-023-00969-x} {\path{doi:10.1140/epja/s10050-023-00969-x}}.
\newline\urlprefix\url{https://doi.org/10.1140/epja/s10050-023-00969-x}

\bibitem{Bliss2020}
J.~Bliss, A.~Arcones, F.~Montes, J.~Pereira, \href{https://link.aps.org/doi/10.1103/PhysRevC.101.055807}{{Nuclear physics uncertainties in neutrino-driven, neutron-rich supernova ejecta}}, Phys. Rev. C 101 (2020) 055807.
\newblock \href {https://doi.org/10.1103/PhysRevC.101.055807} {\path{doi:10.1103/PhysRevC.101.055807}}.
\newline\urlprefix\url{https://link.aps.org/doi/10.1103/PhysRevC.101.055807}

\bibitem{Eronen2012}
T.~Eronen, V.~S. Kolhinen, V.~V. Elomaa, D.~Gorelov, U.~Hager, J.~Hakala, A.~Jokinen, A.~Kankainen, P.~Karvonen, S.~Kopecky, I.~D. Moore, H.~Penttil{\"a}, S.~Rahaman, S.~Rinta-Antila, J.~Rissanen, A.~Saastamoinen, J.~Szerypo, C.~Weber, J.~{\"A}yst{\"o}, \href{https://doi.org/10.1140/epja/i2012-12046-1}{{{JYFLTRAP}: a {P}enning trap for precision mass spectroscopy and isobaric purification}}, The European Physical Journal A 48~(4) (2012) 46.
\newblock \href {https://doi.org/10.1140/epja/i2012-12046-1} {\path{doi:10.1140/epja/i2012-12046-1}}.
\newline\urlprefix\url{https://doi.org/10.1140/epja/i2012-12046-1}

\bibitem{Moore2013}
I.~Moore, T.~Eronen, D.~Gorelov, J.~Hakala, A.~Jokinen, A.~Kankainen, V.~Kolhinen, J.~Koponen, H.~Penttilä, I.~Pohjalainen, M.~Reponen, J.~Rissanen, A.~Saastamoinen, S.~Rinta-Antila, V.~Sonnenschein, J.~Äystö, \href{https://www.sciencedirect.com/science/article/pii/S0168583X13007143}{{Towards commissioning the new {IGISOL}-4 facility}}, Nuclear Instruments and Methods in Physics Research Section B: Beam Interactions with Materials and Atoms 317 (2013) 208--213, {XVI}th {I}nternational {C}onference on {E}lectro{M}agnetic {I}sotope {S}eparators and {T}echniques {R}elated to their {A}pplications, {D}ecember 2–7, 2012 at {M}atsue, {J}apan.
\newblock \href {https://doi.org/10.1016/j.nimb.2013.06.036} {\path{doi:10.1016/j.nimb.2013.06.036}}.
\newline\urlprefix\url{https://www.sciencedirect.com/science/article/pii/S0168583X13007143}

\bibitem{Penttila2020}
H.~Penttil\"a, O.~Beliuskina, L.~Canete, A.~de~Roubin, T.~Eronen, M.~Hukkanen, A.~Kankainen, I.~Moore, D.~Nesterenko, P.~Papadakis, I.~Pohjalainen, M.~Reponen, S.~Rinta-Antila, J.~Sar\'en, J.~Uusitalo, M.~Vil\'en, V.~Virtanen, \href{https://doi.org/10.1051/epjconf/202023917002}{{Radioactive ion beam manipulation at the {IGISOL}-4 facility}}, EPJ Web Conf. 239 (2020) 17002.
\newblock \href {https://doi.org/10.1051/epjconf/202023917002} {\path{doi:10.1051/epjconf/202023917002}}.
\newline\urlprefix\url{https://doi.org/10.1051/epjconf/202023917002}

\bibitem{Karvonen2008}
P.~Karvonen, I.~Moore, T.~Sonoda, T.~Kessler, H.~Penttilä, K.~Peräjärvi, P.~Ronkanen, J.~Äystö, \href{https://www.sciencedirect.com/science/article/pii/S0168583X08009191}{{A sextupole ion beam guide to improve the efficiency and beam quality at {IGISOL}}}, Nuclear Instruments and Methods in Physics Research Section B: Beam Interactions with Materials and Atoms 266~(21) (2008) 4794--4807.
\newblock \href {https://doi.org/10.1016/j.nimb.2008.07.022} {\path{doi:10.1016/j.nimb.2008.07.022}}.
\newline\urlprefix\url{https://www.sciencedirect.com/science/article/pii/S0168583X08009191}

\bibitem{Nieminen2001}
A.~Nieminen, J.~Huikari, A.~Jokinen, J.~Äystö, P.~Campbell, E.~Cochrane, \href{https://www.sciencedirect.com/science/article/pii/S0168900200007506}{Beam cooler for low-energy radioactive ions}, Nuclear Instruments and Methods in Physics Research Section A: Accelerators, Spectrometers, Detectors and Associated Equipment 469~(2) (2001) 244--253.
\newblock \href {https://doi.org/10.1016/S0168-9002(00)00750-6} {\path{doi:10.1016/S0168-9002(00)00750-6}}.
\newline\urlprefix\url{https://www.sciencedirect.com/science/article/pii/S0168900200007506}

\bibitem{Savard1991}
G.~Savard, S.~Becker, G.~Bollen, H.-J. Kluge, R.~Moore, T.~Otto, L.~Schweikhard, H.~Stolzenberg, U.~Wiess, \href{https://www.sciencedirect.com/science/article/pii/0375960191910082}{{A new cooling technique for heavy ions in a {P}enning trap}}, Physics Letters A 158~(5) (1991) 247--252.
\newblock \href {https://doi.org/10.1016/0375-9601(91)91008-2} {\path{doi:10.1016/0375-9601(91)91008-2}}.
\newline\urlprefix\url{https://www.sciencedirect.com/science/article/pii/0375960191910082}

\bibitem{Vilen2020}
M.~Vilén, L.~Canete, B.~Cheal, A.~Giatzoglou, R.~{de Groote}, A.~{de Roubin}, T.~Eronen, S.~Geldhof, A.~Jokinen, A.~Kankainen, I.~Moore, D.~Nesterenko, H.~Penttilä, I.~Pohjalainen, M.~Reponen, S.~Rinta-Antila, \href{https://www.sciencedirect.com/science/article/pii/S0168583X19302344}{{A new off-line ion source facility at IGISOL}}, Nuclear Instruments and Methods in Physics Research Section B: Beam Interactions with Materials and Atoms 463 (2020) 382--383.
\newblock \href {https://doi.org/10.1016/j.nimb.2019.04.051} {\path{doi:10.1016/j.nimb.2019.04.051}}.
\newline\urlprefix\url{https://www.sciencedirect.com/science/article/pii/S0168583X19302344}

\bibitem{Roux2013}
C.~Roux, K.~Blaum, M.~Block, C.~Droese, S.~Eliseev, M.~Goncharov, F.~Herfurth, E.~M. Ramirez, D.~A. Nesterenko, Y.~N. Novikov, L.~Schweikhard, \href{https://doi.org/10.1140/epjd/e2013-40110-x}{{Data analysis of {Q}-value measurements for double-electron capture with {SHIPTRAP}}}, The European Physical Journal D 67~(7) (2013) 146.
\newblock \href {https://doi.org/10.1140/epjd/e2013-40110-x} {\path{doi:10.1140/epjd/e2013-40110-x}}.
\newline\urlprefix\url{https://doi.org/10.1140/epjd/e2013-40110-x}

\bibitem{Nesterenko2021}
D.~A. Nesterenko, T.~Eronen, Z.~Ge, A.~Kankainen, M.~Vilen, \href{https://doi.org/10.1140/epja/s10050-021-00608-3}{{Study of radial motion phase advance during motion excitations in a {P}enning trap and accuracy of {JYFLTRAP} mass spectrometer}}, The European Physical Journal A 57~(11) (2021) 302.
\newblock \href {https://doi.org/10.1140/epja/s10050-021-00608-3} {\path{doi:10.1140/epja/s10050-021-00608-3}}.
\newline\urlprefix\url{https://doi.org/10.1140/epja/s10050-021-00608-3}

\bibitem{NUBASE2020}
F.~Kondev, M.~Wang, W.~Huang, S.~Naimi, G.~Audi, \href{https://doi.org/10.1088/1674-1137/abddae}{{The {NUBASE2020} evaluation of nuclear physics properties}}, Chinese Physics C 45~(3) (2021) 030001.
\newblock \href {https://doi.org/10.1088/1674-1137/abddae} {\path{doi:10.1088/1674-1137/abddae}}.
\newline\urlprefix\url{https://doi.org/10.1088/1674-1137/abddae}

\bibitem{Wohn1978}
F.~K. Wohn, W.~L. Talbert, \href{https://link.aps.org/doi/10.1103/PhysRevC.18.2328}{{Decay energies of gaseous fission products and their daughters for $A=88$ to $93$ and $A=138$ to $142$}}, Phys. Rev. C 18 (1978) 2328--2332.
\newblock \href {https://doi.org/10.1103/PhysRevC.18.2328} {\path{doi:10.1103/PhysRevC.18.2328}}.
\newline\urlprefix\url{https://link.aps.org/doi/10.1103/PhysRevC.18.2328}

\bibitem{Ames1953}
D.~P. Ames, M.~E. Bunker, L.~M. Langer, B.~M. Sorenson, \href{https://link.aps.org/doi/10.1103/PhysRev.91.68}{{The Disintegration of ${\mathrm{Sr}}^{91}$ and ${\mathrm{Y}}^{91m}$}}, Phys. Rev. 91 (1953) 68--74.
\newblock \href {https://doi.org/10.1103/PhysRev.91.68} {\path{doi:10.1103/PhysRev.91.68}}.
\newline\urlprefix\url{https://link.aps.org/doi/10.1103/PhysRev.91.68}

\bibitem{Mach1990}
H.~Mach, E.~K. Warburton, R.~L. Gill, R.~F. Casten, J.~A. Becker, B.~A. Brown, J.~A. Winger, \href{https://link.aps.org/doi/10.1103/PhysRevC.41.226}{{Meson-exchange enhancement of the first-forbidden $^{96}\mathrm{Y}^{\mathrm{g}}$(${0}^{\mathrm{\ensuremath{-}}}$)${\ensuremath{\rightarrow}}^{96}$${\mathrm{Zr}}^{\mathit{g}}$ (${0}^{+}$) \ensuremath{\beta} transition: \ensuremath{\beta} decay of the low-spin isomer of $^{96}\mathrm{Y}$}}, Phys. Rev. C 41 (1990) 226--242.
\newblock \href {https://doi.org/10.1103/PhysRevC.41.226} {\path{doi:10.1103/PhysRevC.41.226}}.
\newline\urlprefix\url{https://link.aps.org/doi/10.1103/PhysRevC.41.226}

\bibitem{Kellerbauer2003}
A.~Kellerbauer, K.~Blaum, G.~Bollen, F.~Herfurth, H.-J. Kluge, M.~Kuckein, E.~Sauvan, C.~Scheidenberger, L.~Schweikhard, \href{https://doi.org/10.1140/epjd/e2002-00222-0}{{From direct to absolute mass measurements: A study of the accuracy of ISOLTRAP}}, The European Physical Journal D 22~(1) (2003) 53--64.
\newblock \href {https://doi.org/10.1140/epjd/e2002-00222-0} {\path{doi:10.1140/epjd/e2002-00222-0}}.
\newline\urlprefix\url{https://doi.org/10.1140/epjd/e2002-00222-0}

\bibitem{Baglin2013}
C.~M. Baglin, \href{https://www.sciencedirect.com/science/article/pii/S0090375213000665}{{Nuclear Data Sheets for A = 91}}, Nuclear Data Sheets 114~(10) (2013) 1293--1495.
\newblock \href {https://doi.org/10.1016/j.nds.2013.10.002} {\path{doi:10.1016/j.nds.2013.10.002}}.
\newline\urlprefix\url{https://www.sciencedirect.com/science/article/pii/S0090375213000665}

\bibitem{Rice2017}
S.~Rice, A.~Algora, J.~L. Tain, E.~Valencia, J.~Agramunt, B.~Rubio, W.~Gelletly, P.~H. Regan, A.-A. Zakari-Issoufou, M.~Fallot, A.~Porta, J.~Rissanen, T.~Eronen, J.~\"Ayst\"o, L.~Batist, M.~Bowry, V.~M. Bui, R.~Caballero-Folch, D.~Cano-Ott, V.-V. Elomaa, E.~Estevez, G.~F. Farrelly, A.~R. Garcia, B.~Gomez-Hornillos, V.~Gorlychev, J.~Hakala, M.~D. Jordan, A.~Jokinen, V.~S. Kolhinen, F.~G. Kondev, T.~Mart\'{\i}nez, P.~Mason, E.~Mendoza, I.~Moore, H.~Penttil\"a, Z.~Podoly\'ak, M.~Reponen, V.~Sonnenschein, A.~A. Sonzogni, P.~Sarriguren, \href{https://link.aps.org/doi/10.1103/PhysRevC.96.014320}{{Total absorption spectroscopy study of the $\ensuremath{\beta}$ decay of $^{86}\mathrm{Br}$ and $^{91}\mathrm{Rb}$}}, Phys. Rev. C 96 (2017) 014320.
\newblock \href {https://doi.org/10.1103/PhysRevC.96.014320} {\path{doi:10.1103/PhysRevC.96.014320}}.
\newline\urlprefix\url{https://link.aps.org/doi/10.1103/PhysRevC.96.014320}

\end{thebibliography}

\end{document}